\pdfoutput=1

\documentclass[sigconf,screen,anonymous=false,review=false,nonacm]{acmart}

\AtBeginDocument{%
  }

\setcopyright{acmcopyright}
\copyrightyear{2023}
\acmYear{2023}
\acmDOI{XXXXXXX.XXXXXXX}

\acmConference[WSDM '24]{17th ACM International Conference on Web Search and Data Mining}{March 4--8, 2024}{Mexico}
\acmPrice{15.00}
\acmISBN{978-1-4503-XXXX-X/18/06}

\usepackage[utf8]{inputenc} %
\usepackage[T1]{fontenc}    %
\usepackage{hyperref}       %
\usepackage{url}            %
\usepackage{booktabs}       %
\usepackage{amsmath}        %
\usepackage{amsfonts}       %
\usepackage{nicefrac}       %
\usepackage{microtype}      %
\usepackage{xcolor}         %
\usepackage{enumitem} 
\usepackage{float}
\usepackage{graphicx}
\usepackage{caption}
\usepackage{subcaption} 
\usepackage{tabularx}
\usepackage{multirow}
\usepackage{algorithm}
\usepackage{algpseudocode}
\usepackage{amssymb}       %
\usepackage{pifont}
\usepackage{textcomp}
\usepackage{soul}

\DeclareMathOperator*{\argmax}{argmax} 

\begin{document}

\title{Talk the Walk: Synthetic Data Generation for Conversational Music Recommendation}

 \settopmatter{authorsperrow=4}

\author{Megan Leszczynski}
\authornote{Core contributor.}
\authornote{Work done while interning at Google Research.}
\affiliation{%
  \institution{Stanford University}
    \country{} %
}
\email{mleszczy@cs.stanford.edu}

\author{Shu Zhang}
\authornotemark[1]
\affiliation{%
  \institution{Google Research}
    \country{}
}
\email{shzhang@google.com}

\author{Ravi Ganti}
\authornotemark[1]
\affiliation{%
  \institution{Google Research}
    \country{}
}
\email{gmravi@google.com}

\author{Krisztian Balog}
\affiliation{%
 \institution{Google Research}
   \country{}
 }
 \email{krisztianb@google.com}

\author{Filip Radlinski}
\affiliation{%
 \institution{Google Research}
   \country{}
 }
\email{filiprad@google.com}

\author{Fernando Pereira}
\affiliation{%
  \institution{Google Research}
\country{}
 }
\email{pereira@google.com}

\author{\mbox{Arun Tejasvi Chaganty}}
\authornotemark[1]
\affiliation{%
  \institution{Google Research}
  \country{}
 }
\email{arunchaganty@google.com}

\renewcommand{\shortauthors}{Leszczynski et al.}

\providecommand\sa{\ensuremath{\mathcal{a}}}
\providecommand\sd{\ensuremath{\mathcal{d}}}
\providecommand\se{\ensuremath{\mathcal{e}}}
\providecommand\sg{\ensuremath{\mathcal{g}}}
\providecommand\sh{\ensuremath{\mathcal{h}}}
\providecommand\si{\ensuremath{\mathcal{i}}}
\providecommand\sj{\ensuremath{\mathcal{j}}}
\providecommand\sk{\ensuremath{\mathcal{k}}}
\providecommand\sm{\ensuremath{\mathcal{m}}}
\providecommand\sn{\ensuremath{\mathcal{n}}}
\providecommand\so{\ensuremath{\mathcal{o}}}
\providecommand\sq{\ensuremath{\mathcal{q}}}
\providecommand\sr{\ensuremath{\mathcal{r}}}
\providecommand\st{\ensuremath{\mathcal{t}}}
\providecommand\su{\ensuremath{\mathcal{u}}}
\providecommand\sv{\ensuremath{\mathcal{v}}}
\providecommand\sw{\ensuremath{\mathcal{w}}}
\providecommand\sx{\ensuremath{\mathcal{x}}}
\providecommand\sy{\ensuremath{\mathcal{y}}}
\providecommand\sz{\ensuremath{\mathcal{z}}}
\providecommand\sA{\ensuremath{\mathcal{A}}}
\providecommand\sB{\ensuremath{\mathcal{B}}}
\providecommand\sC{\ensuremath{\mathcal{C}}}
\providecommand\sD{\ensuremath{\mathcal{D}}}
\providecommand\sE{\ensuremath{\mathcal{E}}}
\providecommand\sF{\ensuremath{\mathcal{F}}}
\providecommand\sG{\ensuremath{\mathcal{G}}}
\providecommand\sH{\ensuremath{\mathcal{H}}}
\providecommand\sI{\ensuremath{\mathcal{I}}}
\providecommand\sJ{\ensuremath{\mathcal{J}}}
\providecommand\sK{\ensuremath{\mathcal{K}}}
\providecommand\sL{\ensuremath{\mathcal{L}}}
\providecommand\sM{\ensuremath{\mathcal{M}}}
\providecommand\sN{\ensuremath{\mathcal{N}}}
\providecommand\sO{\ensuremath{\mathcal{O}}}
\providecommand\sP{\ensuremath{\mathcal{P}}}
\providecommand\sQ{\ensuremath{\mathcal{Q}}}
\providecommand\sR{\ensuremath{\mathcal{R}}}
\providecommand\sS{\ensuremath{\mathcal{S}}}
\providecommand\sT{\ensuremath{\mathcal{T}}}
\providecommand\sU{\ensuremath{\mathcal{U}}}
\providecommand\sV{\ensuremath{\mathcal{V}}}
\providecommand\sW{\ensuremath{\mathcal{W}}}
\providecommand\sX{\ensuremath{\mathcal{X}}}
\providecommand\sY{\ensuremath{\mathcal{Y}}}
\providecommand\sZ{\ensuremath{\mathcal{Z}}}
\providecommand\ba{\ensuremath{\mathbf{a}}}
\providecommand\bb{\ensuremath{\mathbf{b}}}
\providecommand\bc{\ensuremath{\mathbf{c}}}
\providecommand\bd{\ensuremath{\mathbf{d}}}
\providecommand\be{\ensuremath{\mathbf{e}}}
\providecommand\bg{\ensuremath{\mathbf{g}}}
\providecommand\bh{\ensuremath{\mathbf{h}}}
\providecommand\bi{\ensuremath{\mathbf{i}}}
\providecommand\bj{\ensuremath{\mathbf{j}}}
\providecommand\bk{\ensuremath{\mathbf{k}}}
\providecommand\bl{\ensuremath{\mathbf{l}}}
\providecommand\bn{\ensuremath{\mathbf{n}}}
\providecommand\bo{\ensuremath{\mathbf{o}}}
\providecommand\bp{\ensuremath{\mathbf{p}}}
\providecommand\bq{\ensuremath{\mathbf{q}}}
\providecommand\br{\ensuremath{\mathbf{r}}}
\providecommand\bs{\ensuremath{\mathbf{s}}}
\providecommand\bt{\ensuremath{\mathbf{t}}}
\providecommand\bu{\ensuremath{\mathbf{u}}}
\providecommand\bv{\ensuremath{\mathbf{v}}}
\providecommand\bw{\ensuremath{\mathbf{w}}}
\providecommand\bx{\ensuremath{\mathbf{x}}}
\providecommand\by{\ensuremath{\mathbf{y}}}
\providecommand\bz{\ensuremath{\mathbf{z}}}
\providecommand\bA{\ensuremath{\mathbf{A}}}
\providecommand\bB{\ensuremath{\mathbf{B}}}
\providecommand\bC{\ensuremath{\mathbf{C}}}
\providecommand\bD{\ensuremath{\mathbf{D}}}
\providecommand\bE{\ensuremath{\mathbf{E}}}
\providecommand\bF{\ensuremath{\mathbf{F}}}
\providecommand\bG{\ensuremath{\mathbf{G}}}
\providecommand\bH{\ensuremath{\mathbf{H}}}
\providecommand\bI{\ensuremath{\mathbf{I}}}
\providecommand\bJ{\ensuremath{\mathbf{J}}}
\providecommand\bK{\ensuremath{\mathbf{K}}}
\providecommand\bL{\ensuremath{\mathbf{L}}}
\providecommand\bM{\ensuremath{\mathbf{M}}}
\providecommand\bN{\ensuremath{\mathbf{N}}}
\providecommand\bO{\ensuremath{\mathbf{O}}}
\providecommand\bP{\ensuremath{\mathbf{P}}}
\providecommand\bQ{\ensuremath{\mathbf{Q}}}
\providecommand\bR{\ensuremath{\mathbf{R}}}
\providecommand\bS{\ensuremath{\mathbf{S}}}
\providecommand\bT{\ensuremath{\mathbf{T}}}
\providecommand\bU{\ensuremath{\mathbf{U}}}
\providecommand\bV{\ensuremath{\mathbf{V}}}
\providecommand\bW{\ensuremath{\mathbf{W}}}
\providecommand\bX{\ensuremath{\mathbf{X}}}
\providecommand\bY{\ensuremath{\mathbf{Y}}}
\providecommand\bZ{\ensuremath{\mathbf{Z}}}
\providecommand\Ba{\ensuremath{\mathbb{a}}}
\providecommand\Bb{\ensuremath{\mathbb{b}}}
\providecommand\Bc{\ensuremath{\mathbb{c}}}
\providecommand\Bd{\ensuremath{\mathbb{d}}}
\providecommand\Be{\ensuremath{\mathbb{e}}}
\providecommand\Bf{\ensuremath{\mathbb{f}}}
\providecommand\Bg{\ensuremath{\mathbb{g}}}
\providecommand\Bh{\ensuremath{\mathbb{h}}}
\providecommand\Bi{\ensuremath{\mathbb{i}}}
\providecommand\Bj{\ensuremath{\mathbb{j}}}
\providecommand\Bk{\ensuremath{\mathbb{k}}}
\providecommand\Bl{\ensuremath{\mathbb{l}}}
\providecommand\Bm{\ensuremath{\mathbb{m}}}
\providecommand\Bn{\ensuremath{\mathbb{n}}}
\providecommand\Bo{\ensuremath{\mathbb{o}}}
\providecommand\Bp{\ensuremath{\mathbb{p}}}
\providecommand\Bq{\ensuremath{\mathbb{q}}}
\providecommand\Br{\ensuremath{\mathbb{r}}}
\providecommand\Bs{\ensuremath{\mathbb{s}}}
\providecommand\Bt{\ensuremath{\mathbb{t}}}
\providecommand\Bu{\ensuremath{\mathbb{u}}}
\providecommand\Bv{\ensuremath{\mathbb{v}}}
\providecommand\Bw{\ensuremath{\mathbb{w}}}
\providecommand\Bx{\ensuremath{\mathbb{x}}}
\providecommand\By{\ensuremath{\mathbb{y}}}
\providecommand\Bz{\ensuremath{\mathbb{z}}}
\providecommand\BA{\ensuremath{\mathbb{A}}}
\providecommand\BB{\ensuremath{\mathbb{B}}}
\providecommand\BC{\ensuremath{\mathbb{C}}}
\providecommand\BD{\ensuremath{\mathbb{D}}}
\providecommand\BE{\ensuremath{\mathbb{E}}}
\providecommand\BF{\ensuremath{\mathbb{F}}}
\providecommand\BG{\ensuremath{\mathbb{G}}}
\providecommand\BH{\ensuremath{\mathbb{H}}}
\providecommand\BI{\ensuremath{\mathbb{I}}}
\providecommand\BJ{\ensuremath{\mathbb{J}}}
\providecommand\BK{\ensuremath{\mathbb{K}}}
\providecommand\BL{\ensuremath{\mathbb{L}}}
\providecommand\BM{\ensuremath{\mathbb{M}}}
\providecommand\BN{\ensuremath{\mathbb{N}}}
\providecommand\BO{\ensuremath{\mathbb{O}}}
\providecommand\BP{\ensuremath{\mathbb{P}}}
\providecommand\BQ{\ensuremath{\mathbb{Q}}}
\providecommand\BR{\ensuremath{\mathbb{R}}}
\providecommand\BS{\ensuremath{\mathbb{S}}}
\providecommand\BT{\ensuremath{\mathbb{T}}}
\providecommand\BU{\ensuremath{\mathbb{U}}}
\providecommand\BV{\ensuremath{\mathbb{V}}}
\providecommand\BW{\ensuremath{\mathbb{W}}}
\providecommand\BX{\ensuremath{\mathbb{X}}}
\providecommand\BY{\ensuremath{\mathbb{Y}}}
\providecommand\BZ{\ensuremath{\mathbb{Z}}}
\providecommand\balpha{\ensuremath{\mbox{\boldmath$\alpha$}}}
\providecommand\bbeta{\ensuremath{\mbox{\boldmath$\beta$}}}
\providecommand\btheta{\ensuremath{\mbox{\boldmath$\theta$}}}
\providecommand\bphi{\ensuremath{\mbox{\boldmath$\phi$}}}
\providecommand\bpi{\ensuremath{\mbox{\boldmath$\pi$}}}
\providecommand\bpsi{\ensuremath{\mbox{\boldmath$\psi$}}}
\providecommand\bmu{\ensuremath{\mbox{\boldmath$\mu$}}}
\newcommand\fig[1]{\begin{center} \includegraphics{#1} \end{center}}
\newcommand\Fig[4]{\begin{figure}[ht] \begin{center} \includegraphics[scale=#2]{#1} \end{center} \caption{\label{fig:#3} #4} \end{figure}}
\newcommand\FigTop[4]{\begin{figure}[t] \begin{center} \includegraphics[scale=#2]{#1} \end{center} \caption{\label{fig:#3} #4} \end{figure}}
\newcommand\FigStar[4]{\begin{figure*}[ht] \begin{center} \includegraphics[scale=#2]{#1} \end{center} \caption{\label{fig:#3} #4} \end{figure*}}
\newcommand\FigRight[4]{\begin{wrapfigure}{r}{0.5\textwidth} \begin{center} \includegraphics[width=0.5\textwidth]{#1} \end{center} \caption{\label{fig:#3} #4} \end{wrapfigure}}
\newcommand\aside[1]{\quad\text{[#1]}}
\newcommand\interpret[1]{\llbracket #1 \rrbracket} %
\newcommand{\var}{\text{Var}} %
\newcommand{\cov}{\text{Cov}} %
\newcommand\p[1]{\ensuremath{\left( #1 \right)}} %
\newcommand\pa[1]{\ensuremath{\left\langle #1 \right\rangle}} %
\newcommand\pb[1]{\ensuremath{\left[ #1 \right]}} %
\newcommand\pc[1]{\ensuremath{\left\{ #1 \right\}}} %
\newcommand\eval[2]{\ensuremath{\left. #1 \right|_{#2}}} %
\newcommand\inv[1]{\ensuremath{\frac{1}{#1}}}
\newcommand\half{\ensuremath{\frac{1}{2}}}
\newcommand\R{\ensuremath{\mathbb{R}}} %
\newcommand\Z{\ensuremath{\mathbb{Z}}} %
\newcommand\inner[2]{\ensuremath{\left< #1, #2 \right>}} %
\newcommand\mat[2]{\ensuremath{\left(\begin{array}{#1}#2\end{array}\right)}} %
\newcommand\eqn[1]{\begin{align} #1 \end{align}} %
\newcommand\eqnl[2]{\begin{align} \label{eqn:#1} #2 \end{align}} %
\newcommand\eqdef{\ensuremath{\stackrel{\rm def}{=}}} %
\newcommand{\1}{\mathbb{I}} %
\newcommand{\bone}{\mathbf{1}} %
\newcommand{\bzero}{\mathbf{0}} %
\newcommand\refeqn[1]{(\ref{eqn:#1})}
\newcommand\refeqns[2]{(\ref{eqn:#1}) and (\ref{eqn:#2})}
\newcommand\refchp[1]{Chapter~\ref{chp:#1}}
\newcommand\refchap[1]{Chapter~\ref{chap:#1}}
\newcommand\refsec[1]{Section~\ref{sec:#1}}
\newcommand\refsecs[2]{Sections~\ref{sec:#1} and~\ref{sec:#2}}
\newcommand\reffig[1]{Figure~\ref{fig:#1}}
\newcommand\reffigs[2]{Figures~\ref{fig:#1} and~\ref{fig:#2}}
\newcommand\reffigss[3]{Figures~\ref{fig:#1},~\ref{fig:#2}, and~\ref{fig:#3}}
\newcommand\reffigsss[4]{Figures~\ref{fig:#1},~\ref{fig:#2},~\ref{fig:#3}, and~\ref{fig:#4}}
\newcommand\reftab[1]{Table~\ref{tab:#1}}
\newcommand\refapp[1]{Appendix~\ref{sec:#1}}
\newcommand\refthm[1]{Theorem~\ref{thm:#1}}
\newcommand\refthms[2]{Theorems~\ref{thm:#1} and~\ref{thm:#2}}
\newcommand\reflem[1]{Lemma~\ref{lem:#1}}
\newcommand\reflems[2]{Lemmas~\ref{lem:#1} and~\ref{lem:#2}}
\newcommand\refalg[1]{Algorithm~\ref{alg:#1}}
\newcommand\refalgs[2]{Algorithms~\ref{alg:#1} and~\ref{alg:#2}}
\newcommand\refex[1]{Example~\ref{ex:#1}}
\newcommand\refexs[2]{Examples~\ref{ex:#1} and~\ref{ex:#2}}
\newcommand\refprop[1]{Proposition~\ref{prop:#1}}
\newcommand\refdef[1]{Definition~\ref{def:#1}}
\newcommand\refcor[1]{Corollary~\ref{cor:#1}}
\newcommand\Chapter[2]{\chapter{#2}\label{chp:#1}}
\newcommand\Section[2]{\section{#2}\label{sec:#1}}
\newcommand\Subsection[2]{\subsection{#2}\label{sec:#1}}
\newcommand\Subsubsection[2]{\subsubsection{#2}\label{sec:#1}}
\ifthenelse{\isundefined{\definition}}{\newtheorem{definition}{Definition}}{}
\ifthenelse{\isundefined{\assumption}}{\newtheorem{assumption}{Assumption}}{}
\ifthenelse{\isundefined{\hypothesis}}{\newtheorem{hypothesis}{Hypothesis}}{}
\ifthenelse{\isundefined{\proposition}}{\newtheorem{proposition}{Proposition}}{}
\ifthenelse{\isundefined{\theorem}}{\newtheorem{theorem}{Theorem}}{}
\ifthenelse{\isundefined{\lemma}}{\newtheorem{lemma}{Lemma}}{}
\ifthenelse{\isundefined{\corollary}}{\newtheorem{corollary}{Corollary}}{}
\ifthenelse{\isundefined{\alg}}{\newtheorem{alg}{Algorithm}}{}
\ifthenelse{\isundefined{\example}}{\newtheorem{example}{Example}}{}
\newcommand\cv{\ensuremath{\to}} %
\newcommand\cvL{\ensuremath{\xrightarrow{\mathcal{L}}}} %
\newcommand\cvd{\ensuremath{\xrightarrow{d}}} %
\newcommand\cvP{\ensuremath{\xrightarrow{P}}} %
\newcommand\cvas{\ensuremath{\xrightarrow{a.s.}}} %
\newcommand\eqdistrib{\ensuremath{\stackrel{d}{=}}} %
\newcommand{\E}{\ensuremath{\mathbb{E}}} %
\newcommand\KL[2]{\ensuremath{\text{KL}\left( #1 \| #2 \right)}} %

\newcommand{\editmode}{}

\ifundef{\editmode}{
\newcommand{\ac}[1]{}
\newcommand{\ml}[1]{}
\newcommand{\rg}[1]{}
\newcommand{\fr}[1]{}
\newcommand{\kb}[1]{}
\newcommand{\fp}[1]{}
\newcommand{\TODO}[1]{}
\newcommand{\edit}[2]{#1}
\newcommand{\revised}[1]{#1}
}{%
\newcommand{\ac}[1]{{\color{olive}{\bf{AC:}} \emph{#1}}}
\newcommand{\ml}[1]{{\color{olive}{\bf{ML:}} \emph{#1}}}
\newcommand{\rg}[1]{{\color{olive}{\bf{RG:}} \emph{#1}}}
\newcommand{\fr}[1]{{\color{olive}{\bf{FR:}} \emph{#1}}}
\newcommand{\kb}[1]{{\color{olive}{\bf{KB:}} \emph{#1}}}
\newcommand{\fp}[1]{{\color{olive}{\bf{FP:}} \emph{#1}}}
\newcommand{\TODO}[1]{{\color{magenta}{\bf{TODO}} \emph{#1}}}
\newcommand{\edit}[2]{{\color{blue} #1}{\st{#2}}}
\newcommand{\revised}[1]{{\color{blue} #1}}
}

\newcommand{\Xstar}{X_{*}}
\newcommand{\bbR}{\mathbb{R}}

\newcommand{\ourdata}{TtWMusic}
\newcommand{\playlistdata}{ExpertPlaylists} 
\newcommand{\UserTemplates}{\texttt{UserTemplates}}

\newcommand{\ourmethod}{TalkTheWalk}
\newcommand{\shortourmethod}{TtW}
\newcommand{\sysname}{DE$\rhd$\ourdata{}}
\newcommand{\sysnamecr}{Cr$\rhd$\ourdata{}}
\newcommand{\shortsysname}{$\rhd$\ourdata{}}
\newcommand{\oursystem}{\sysname{}}
\newcommand{\oursystemcr}{\sysnamecr{}}
\newcommand{\shortoursystem}{\shortsysname{}}
\newcommand{\nonconvo}{DE$\rhd$\playlistdata{}}
\newcommand{\nonconvoDE}{item collection DE}

\newcommand{\mR}{R}

\newcommand{\xr}{\rho}
\newcommand{\xE}{P}

\newcommand{\sep}{\texttt{[SEP]}\ }

\newcommand{\xe}{\xr}

\newcommand{\xu}{u}

\newcommand{\xx}{x}
\newcommand{\xs}{\bs}
\newcommand{\xp}{\bz}
\newcommand{\xc}{\br}
\newcommand{\xxv}{\tilde{\xx}}
\newcommand{\xpv}{\tilde{\xp}}
\newcommand{\xsv}{\tilde{\xs}}
\newcommand{\xcv}{\tilde{\xc}}

\newcommand{\cp} {conversational retrieval~}

\newcommand{\embed}{\operatorname{\textrm{embed}}}
\newcommand*{\defeq}{\stackrel{\text{def}}{=}}

\newcommand{\todo}[1]{{\color{red} #1}}

\newcommand{\revision}[2]{\textst{#1} {\color{blue} #2}}
\newcommand{\edited}[1]{{\color{red} #1}}

\newcommand{\desc}{\textbf{\textit{\$description}}}

\newcommand{\cmark}{\ding{51}}%
\newcommand{\xmark}{\ding{55}}%
\newcommand{\music}[1]{\href{https://music.youtube.com/watch?v=#1}{\textmusicalnote}}%

\begin{abstract}

Recommender systems are ubiquitous yet often difficult for users to control, and adjust if recommendation quality is poor. This has motivated conversational recommender systems (CRSs), with control provided through natural language feedback.
However, as with most application domains, building robust CRSs requires training data that reflects system usage---here conversations with user utterances paired with items that cover a wide range of preferences.
This has proved challenging to collect scalably using conventional methods. 
We address the question of whether it can be generated synthetically, building on recent advances in natural language. We evaluate in the setting of item set recommendation, noting the increasing attention to this task motivated by use cases like music, news, and recipe recommendation.
We present \ourmethod{}, which synthesizes realistic high-quality conversational data by leveraging domain expertise encoded in widely available curated item collections, %
generating a sequence of hypothetical yet plausible item sets, %
then using a language model to produce corresponding user utterances.
We generate over one million diverse playlist curation conversations in the music domain, and show these contain
\textit{consistent} utterances with \textit{relevant} item sets nearly matching the quality of an existing but small human-collected dataset for this task. 
We demonstrate the utility of the generated synthetic dataset on a conversational item retrieval task and show that it improves over both unsupervised baselines and systems trained on a real dataset.

\end{abstract}

\maketitle

\Section{intro}{Introduction}
\begin{figure}[t]
    \centering
    \includegraphics[width=7cm,trim=0 2.5cm 0 0]{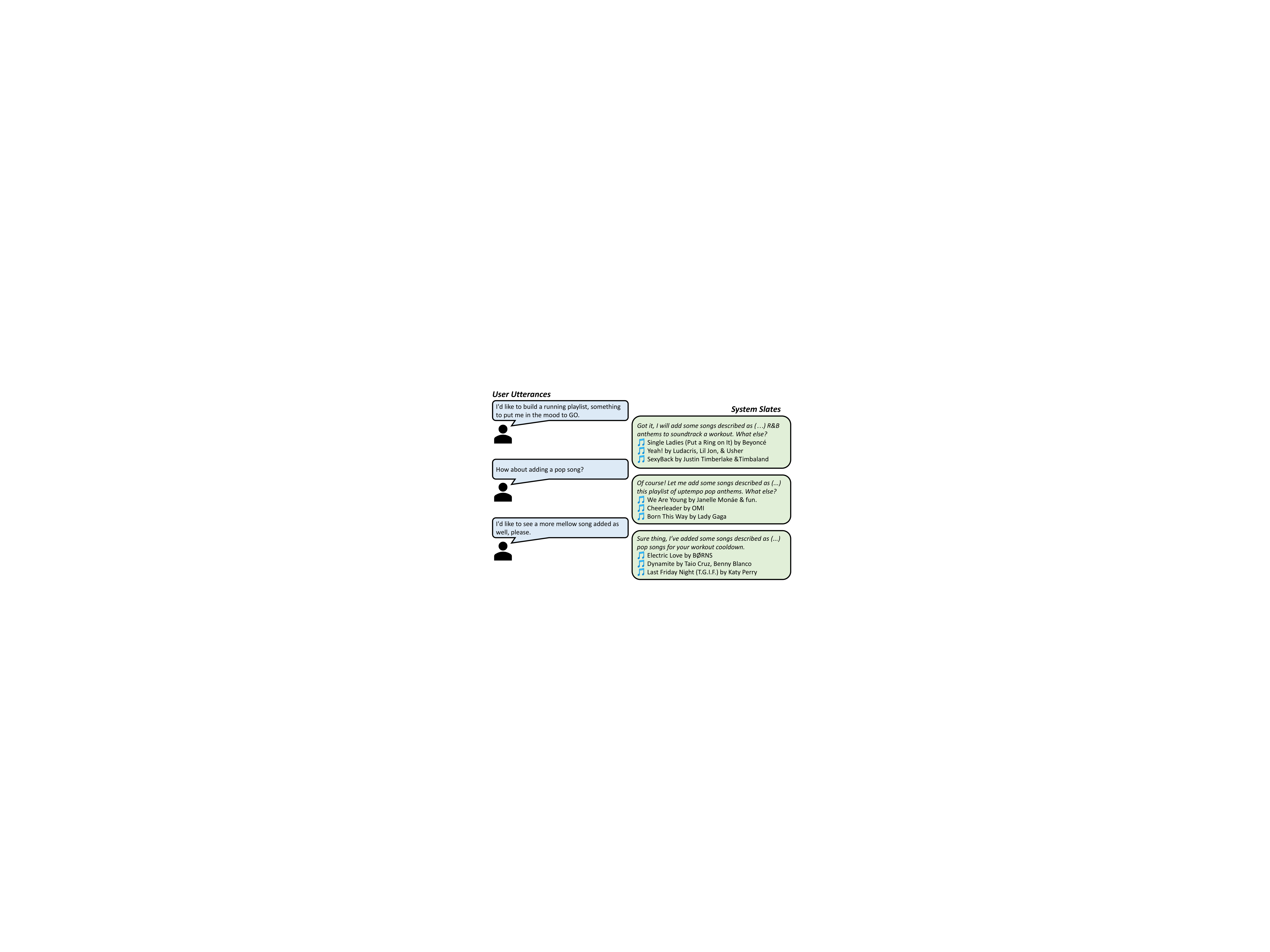}
    \caption{
    \label{fig:overview}
    Example synthetic playlist curation dialog from \ourmethod{} (shortened for presentation). Our domain-agnostic approach generates both conversation sides.
    }
    \vspace{-3mm}
\end{figure}

Recommender systems (RSs) help users choose from an overwhelming number of options. %
Traditionally, they tend to rely on historical interaction data like logs and ratings to personalize results. %
Yet it is well recognized that this gives users limited control over recommendations based on their current context or when preferences change, and also hampers recommendation quality when limited historical data is available in cold-start settings~\cite{jannach-convrec-survey}.  
Recent advances in natural language (NL) processing have enabled \emph{conversational recommender systems} (CRSs) as a step towards addressing many of these challenges, allowing users to engage in a back-and-forth with a CRS~\cite{gao-convrec-survey,jannach-convrec-survey}.  
Specifically, NL feedback allows users to steer the system (e.g., ``How about some more upbeat music?''), reducing the need for historical data while providing current and contextual recommendations. %
While significant progress has been made (see overview \cite{cis-survey}, and more recent advances in language modeling~\citep{Geng:2022:P5,gao2023chatrec}), practical adoption is still limited.
When it comes to the underlying user task, most recommenders primarily assist users find specific items to purchase or consume. While they often facilitate exploration, the ultimate goal is identifying a single item that meets the user's needs and preferences.
We address a different and largely unexplored task: aiding the user via a natural language conversational interface to curate a \textit{set} of items satisfying particular %
preferences, e.g., creating a music playlist for a long drive~\citep{chaganty2023cpcd}.

A key challenge that in the conversational recommendation space is the limited availability of data with natural, diverse and coherent multi-turn conversations paired with relevant per-turn recommendations---meaning CRSs are unable to deliver the full control they should in principle support.
Previous work derives data from actual user-system interactions~\cite{christakopoulou2018q}, or mocked-up human-human ``Wizard of Oz'' interactions~\cite{li2018towards,liu-etal-2020-towards-conversational,moon-etal-2019-opendialkg,hayati-etal-2020-inspired}.
In the former, data is biased by existing systems' limited capabilities; in the latter, by the difficulty in crowdworkers producing realistic, varied and on-topic conversations
at reasonable cost%
~\cite{woz-orig-1984,wen2016network,budzianowski2018multiwoz,li2018towards}. 

Given the challenges in collecting real recommendation conversations, %
we instead show how to generate \emph{high quality} synthetic conversations using non-conversational artifacts already frequently curated. %
Specifically, we observe that user generated content in the form of item collections such as playlists, movie watchlists, or recipe books, are widely available and capture the diverse domain expertise of their creators, 
include coherent sets of items with titles, descriptions, %
and other %
attributes often used to express in preferences (e.g., ``upbeat music,'' ``feel-good movies,'' ``healthy recipes'').
However, item collections lack two essential pieces: (1) multiple turns of user preference descriptions, and (2) corresponding recommended items. 
We %
tackle these in reverse order: generate slates (i.e., sets of recommended items~\cite{kale-nonstochastic}), then generate corresponding utterances. %

Concretely,
we represent items and item collections in an embedding space using a standard dual encoder~\cite{wu2019zero,gillick-etal-2019-learning}.
We then perform a biased random walk in this space to pick a consistent sequence of slates from the item collections.
We use the metadata from the collections to prompt a dialog inpainting language model~\citep{pmlr-v162-dai22a} to generate conversational utterances that express preferences for each slate.
The approach is scalable (like historical interaction data), respects user privacy  (like crowdsourced datasets), and generates representative conversations allowing 
effective CRS training.

We use our approach to create a dataset of over one million synthetic playlist curation conversations covering very diverse domains, from Japanese pop to electro-swing music. 
Music recommendation is an excellent example where conversational recommendation offers huge potential, as long playlist are often incrementally refined by  users reacting to what they are (or are not) hearing.

We evaluate the generated synthetic conversations in terms of \emph{quality} and \emph{utility}.
Qualitative assessment is performed by comparing synthetic conversations against human-collected dialogues~\citep{chaganty2023cpcd} in terms of consistency, relevance, and naturalness. 
Our results show that over 70\% of generated conversations are rated realistic, and that utterance consistency and slate relevance is comparable to a human-collected dataset.
For measuring utility, we leverage synthetic data to train models for the task of conversational item recommendation and show that models trained only on synthetic data or on the combination of real and synthetic data outperform respective baselines. 
Moreover, we estimate that creating it costs less than \$1,000, making it a cost-effective alternative to crowdsourcing.

In summary: 
\begin{itemize}[leftmargin=*,topsep=2pt]
\item We introduce \ourmethod{} (\shortourmethod), a novel method to generate training data for CRSs that leverages curated item collections and demonstrate it produces realistic item set curation conversations. 
\item We apply \shortourmethod{} on expert-curated music playlists, resulting in a synthetic dataset of over one million conversations that is validated against a human-collected dataset. 
\item We show that this data enables training a music CRS that outperforms baselines trained without using synthetic data.%
\end{itemize} 

\section{Related work}

As we address synthetic data generation in a conversational setting for item set recommendation, we review related work in these areas.

\subsubsection*{\bf Synthetic Data Generation}

Synthetic data is frequently used to help address both \emph{privacy} and \emph{data scarcity} challenges for RSs~\cite{provalov-synevarec,stavinova-synthetic-survey}. 
Traditional RSs rely on historical data (i.e., a user-item matrix), leading researchers to note the need to consider user privacy in developing reproducible evaluations~\cite{ramakrishnan-privacy, calandrino-privacy-cf,Jeckmans2013}.
Solutions proposed %
include transforming historical data by partial replacement~\cite{slokom-synthetic,liu-privacy-preserving} or randomized perturbation~\cite{polat-privacy-preserving}.
Our approach avoids historical data. %
At the same time, real user data is often scarce in \textit{new} applications because existing systems are limited in their capability to respond to user requests. Most commonly, previous
approaches generate synthetic data by sampling an explicit probabilistic model; it can be manually configured or made to match aggregate statistics of historical data~\cite{del-carmen-datagencars,pasinato-context-aware}.
However, such approaches only generate attributes and item ratings to express user preferences, rather than natural language utterances as we do.

Closest to our work are studies that generate synthetic data for CRSs using item ratings~\cite{Dodge2015EvaluatingPQ}, text reviews~\cite{zhang-saur}, and system logs~\cite{zhou-etal-2020-towards}.
\citet{Dodge2015EvaluatingPQ} and \citet{zhang-saur} use templates to generate utterances, while \citet{zhou-etal-2020-towards} retrieve candidate utterances then ask human annotators to rewrite them.
In contrast, we use a language model (LM)~\cite{pmlr-v162-dai22a} to generate user utterances.
\citet{zamani-synthetic-data} highlight challenges in evaluating synthetic data generated by large LMs; we evaluate it both directly through a crowdsourcing and indirectly through the performance of models trained on the data. 
Finally, recent work on user simulation~\cite{zhang-user-sim,balog-user-sim} studies how CRS evaluation can be automated, by mimicking how a user would respond at each turn. In contrast, we focus on automating the generation of entire conversations, including queries and slates.%

\subsubsection*{\bf Conversational Recommendation}
CRSs %
help users find items of interest through a sequence of interactions (i.e., conversation)~\cite{jannach-convrec-survey}. 
It often encompasses recommendation, conversational search, and conversational question answering~\cite{cis-survey}. 
They invite users to directly respond to improve recommendations, contrasting with historical interaction data-dominated collaborative filtering-based RSs~\citep{koren2009matrix, koren2022advances}. 
However limited
conversational training data means
CRSs are often trained using reinforcement learning~\citep{zhao2013interactive, zou2020neural, christakopoulou2016towards, lei2020estimation}, or supervised learning on scripted dialogues~\citep{iai_moviebot}.
Recent work often collects conversational data through crowdworkers~\cite{li2018towards,zhou-etal-2020-towards,kang-etal-2019-recommendation,moon-etal-2019-opendialkg,liu-etal-2020-towards-conversational,hayati-etal-2020-inspired,liu-etal-2021-durecdial,taskmaster2,chaganty2023cpcd}, usually limited to domains like movie recommendation, and relatively small sizes (e.g., the popular ReDial dataset~\cite{li2018towards} has only $\sim$10k conversations), which can promote overfitting~\cite{wang-etal-2022-recindial}.
While for individual item recommendations, a CRS can ask users about preferences on specific attributes~\citep{zhang-saur,sun2018conversational}, item sets like playlists cannot easily be characterized as simply, rather often needing gradual refinement to produce a balanced collection.
Closely related to CRSs is example critiquing, where users can provide feedback on slates, using a fixed set of attributes or tags (e.g., ``more upbeat'')~\citep{cav-recsys-2022}. 
\citet{cav-recsys-2022} build a RS by modeling critiques as updates to a user embedding with learned concept activation vectors (CAVs), which represent the tags (e.g., ``upbeat'') in an embedding space.  
We apply similar techniques to synthetically generate training data, but use item collection embeddings, rather than a fixed set of CAVs, to approximate preferences.
Our work also builds on work in conversational search and conversational question answering~\cite{yu2021convdr,krasakis-zeroshot,mao-curriculum}.

In the context of recent progress on large language models, this new form of conversational recommendation has rapidly attracted attention~\cite{wu2023survey}. Our focus complements these advancements by showing how to create quantifiably realistic recommendation conversations synthetically. Further, we show how this data allows a competitive CRS supporting natural-language to be built on top of state-of-the-art dual encoder recommendation approaches.

\subsubsection*{\bf Item Set Recommendation}

Traditionally, recommender systems help users find  specific items to buy or consume. 
Recently, several real-life scenarios have been identified where recommendations consist of sets of items that need to be considered together, e.g., outfits~\citep{Chen:2019:KDD} or shopping baskets~\citep{Wan:2018:CIKM}, highlighting the importance of set recommendation.
Music playlists have also been studied, %
aiming to ``create a sequence of tracks fulfilling the target characteristics in the best possible way''~\citep{Bonnin:2014:CS}.
Here, we do not focus on the sequence of tracks, but rather on aiding the user identify items to include in a playlist.
Viewing playlists as sets also relates to entity list completion (or \emph{example-augmented search}) %
\citep{Balog:2018:EOS}. There, input consists of a text query and a small set of examples. Benchmarks addressing this problem include INEX 2007--2009 Entity Ranking~\citep{Demartini:2010:OIE} and the TREC 2010--2011 Entity track~\citep{balog12trecentity}. %
We also follow recent work in dense entity retrieval, adopting a dual encoder to embed queries and items~\cite{gillick-etal-2019-learning,wu2019zero,leszczynski-etal-2022-tabi}. 
Novel to our work is that item sets are selected with explicit user feedback, %
in contrast to %
implicit signals of music streaming services through actions like %
skipping songs~\citep{Garcia-Gathright:2018:SIGIR}.

The \emph{slate recommendation} task is also related: It involves grouping together and presenting items sets %
called \emph{slates}. Slate generation approaches include List Conditional Variational Auto-Encoders~\citep{jiang2018beyond} and Slate-MDPs~\citep{sunehag2015deep}. \citet{Xian:2021:Item-Set} extend slate generation to include explanations based on important item attributes.
Unlike those, we generate slates based on natural language conversations.

\Section{method}{Synthetic Item Set Curation Conversation Generation}

\begin{figure*}
    \begin{center}
        \includegraphics[width=0.92\textwidth,trim=0 2cm 0 0]{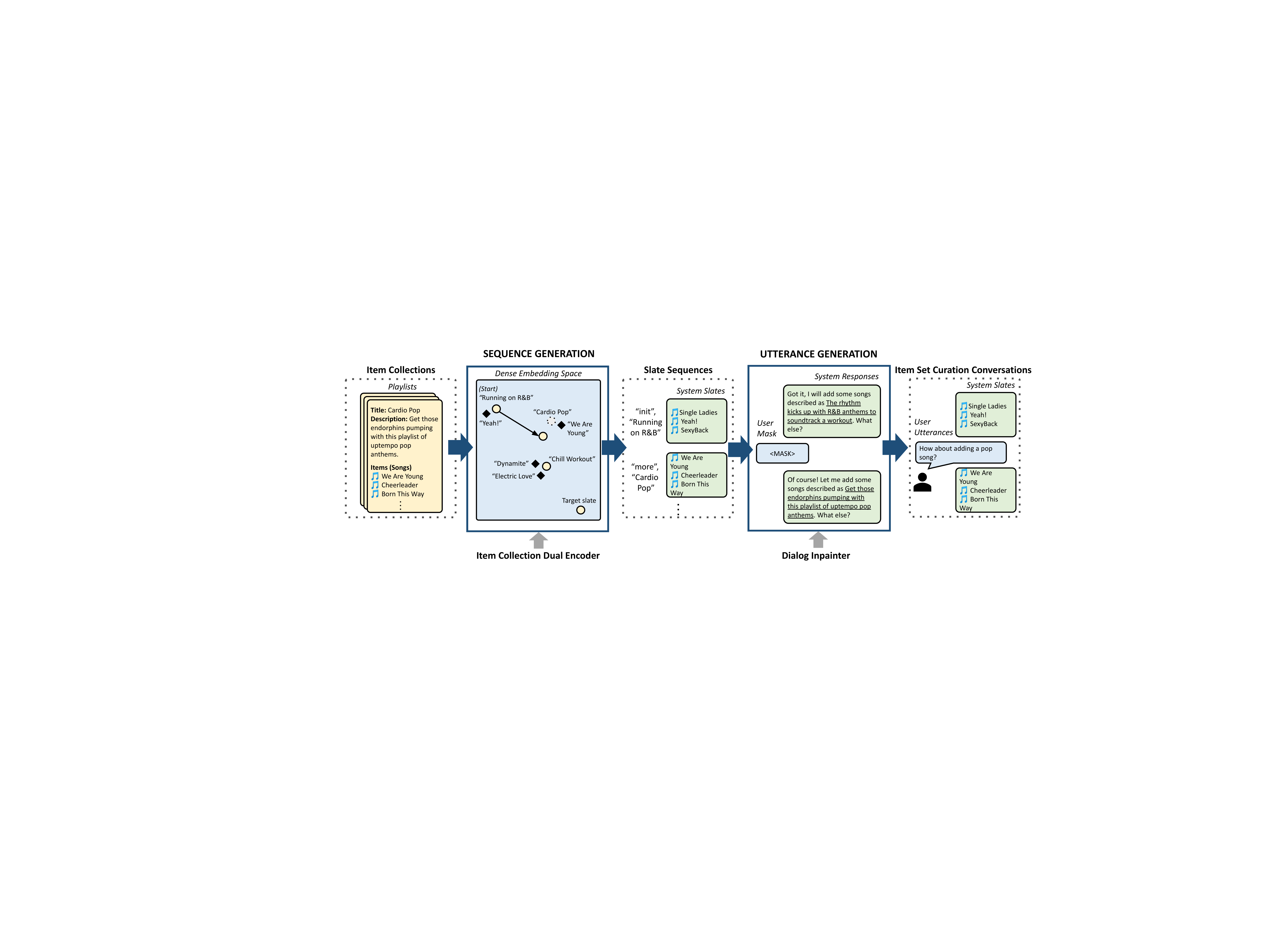}
    \end{center}
    \caption{\label{fig:approach}
        \ourmethod{}: %
        First, sequence generation creates slate sequences using a random walk in an embedding space.
        Then, utterances are generated by prompting a dialog inpainter LM with templated system responses using slate metadata. This %
        gives us complete item set curation conversations. %
        Note the templated system responses are not used after utterance generation.%
        }
\end{figure*}

We now present our main contribution: The \ourmethod{} method to generate synthetic conversational data that is consistent, coherent, makes conversational progress, and is natural. 
This builds on user generated content, specifically here on curated item collections that exist in many application domains---with significant domain expertise by grouping items into coherent collections with rich metadata. Our method also takes advantage of existing language and embedding models.

Existing curated collections lack two necessary features: %
a sequence of user utterances describing preferences (instead of non-conversational metadata), and a corresponding sequence of item slates presented in responses %
(instead of fixed collections).
Thus key innovations include using item collections to generate sequences of slates, then generating user utterances for the sequences.

We now discuss the input and output in detail, and the two main components: sequence generation and utterance generation. Finally, we apply \ourmethod{} to 
to music, taking playlists as the curated collections. %
We generate a new dataset with over 1M synthetic playlist curation conversations that we use to validate the approach in the later evaluation sections.
See \reffig{approach} for an overview.%

\Subsection{system-inputs}{Input and Output} 

\subsubsection*{\bf Input} Assume a corpus of items $\xx \in \sX$ (e.g., songs in a music corpus). 
The main input to data generation is item collections $\sZ$.
Each $\xp \in \sZ$ is a set of items $x\in\sX$ sharing coherent metadata $\phi(\xp)$, for instance a playlist and its description.\footnote{For the scope of this paper, we ignore the ordering of items in a collection, which can be important, e.g., to convey a specific mood or chronology.}
Item collections can be categorized by type $\sigma(\xp) \in \sT$ (e.g., mood or artist-based playlists).
In addition, we assume two pretrained models as input:
a pretrained dual encoder to represent item collections and items in a shared dense embedding space (hereafter referred to as the \nonconvoDE)---which already exist for existing recommendation tasks;
and a pretrained \textit{dialog inpainter}, a language model approach that predicts masked conversation utterances~\cite{pmlr-v162-dai22a}. Formally, given a partial $T$ turn conversation $\bC_T = \left(u_1,...,u_{t-1}, \Diamond, u_{t+1},...,u_T\right)$, where $\Diamond$ represents a masked utterance, a dialog inpainter specifies a probability distribution $p_\theta\left(u_t \mid \bC_T\right)$ parameterized by  $\theta$. 

\subsubsection*{\bf Output} The output of data generation is an item set curation conversation with $T$ turns $\bC_T = \left(u_1,\xs_1, ..., u_T, \xs_T\right)$, where each turn $t$ has a user utterance $u_t$ and a corresponding slate of items $\xs_t \subset \sX$.

\Subsection{seq-gen}{Sequence Generation}

Aiming to generate a realistic sequence of slates that may occur in a conversation, we discuss desired properties, then present our approach using item collections %
(see \reffig{approach} left and \refalg{seq-gen}). 

\subsubsection{Desired Properties} We are guided by following desirable properties for slates in an ideal conversation:
\begin{enumerate}[leftmargin=*, label=({P.}{{\arabic*}})] 
\item Slate $\xs_t$ should be \textit{consistent} with, or closely related to, the previous turn's slate $\xs_{t-1}$. 
If the user first asked for workout music, they are more likely to ask for pop %
than classical music.
\item The change between consecutive slates should be \textit{coherent}, corresponding to user preferences expressed in natural language; %
item collection $\xp_t$ should approximate this change.
\item Each turn should bring the user \textit{closer to their target slate} $\xs^*$. If the user ultimately wants good workout music, we expect slates to include more high energy music in later turns.\footnote{%
Real conversations can also ``backtrack,'' where users revert earlier preferences. While out of scope, such behavior could be modelled resampling $\xsv_t$ and $s^*$.}
\end{enumerate}

\subsubsection{Approach}
To realize these properties, we start by representing items $\xx$, item collections $\xp_t$, and slates $\xs_t$ as vectors ($\xxv$, $\xpv_{t}$, $\xsv_{t}$) in a shared embedding space $\R^d$ using 
the \nonconvoDE.

To maintain consistency across turns %
we represent the preferences at turn $t$ by vector
$\xcv_t$ and ensure that $\xcv_{t+1}$ is close to $\xcv_{t}$ and $\xsv_t$ using cosine distance.
We also assume that each user has a target vector $\xcv^*$ representing their target slate $\xs^*$. We sample the initial user vector $\xcv_1$ and the target user vector $\xcv^*$ from $\sZ$ to ensure they are coherent (i.e., represent real item collections).\footnote{Specifically, we first sample $\xcv^*$ from $\sZ$ and then sample $\xcv_1$ from the item collection neighbors of $\xcv^*$ with a neighbor index in [64, 128) to encourage $\xcv_1, \xcv^*$ to be related.} 
To make the changes between consecutive slates coherent, we model each update as a linear combination of the user vector $\xcv_{t}$ and a nearby item collection vector $\xpv_{t}$: 
\begin{align}
\label{eqn:linear-combo}
\xcv_{t+1} = \alpha \xcv_{t} + \beta \xpv_{t}
\end{align}
\noindent where $\alpha$ and $\beta$ are $t$-dependent parameters described below. To select $\xpv_{t}$, we randomly sample an item collection type $\sigma_t \in \sT$, then 
sample an item collection of that type from the neighborhood of $\xcv_{t}$.\footnote{Based on preliminary experiments, we use softmax($\sN(\xcv_{t})\xcv^*/\tau$) as the sampling distribution (neighborhood size 64 and softmax temperature $\tau=0.1$).}
Intuitively, $\beta > 0$ means a positive preference for $\xp_t$ and moves the user vector in the direction of $\xpv_t$. $\beta < 0$ represents the opposite, moving the user vector in the opposite direction.
The item collection type corresponds to different types of user preferences, e.g.~a mood preference or artist preference when looking for music.

Finally, to ensure each turn brings the user closer to their target slate, we choose the weights $\alpha$ and $\beta$ that minimize the distance to $\xcv^*$ while keeping $\xcv_{t+1}$ unit norm:
\begin{align}
\alpha^*, \beta^* = \argmax_{\alpha, \beta}\ \langle\alpha \xcv_{t} + \beta\xpv_{t}, \xcv^*\rangle\ 
\textrm{s.t.}\ \| \alpha \xcv_{t} + \beta\xpv_{t}\|_2 = 1. \label{eqn:opt-weights}
\end{align}
Optimal values of $\alpha$ and $\beta$ can be found
in closed form 
as one of:
\begin{align}
\alpha &=
  \frac{\pm (w-qv)}
  {\sqrt{(q^2 - 1)^2v^2 - (w-qv)^2(q^2-1)}} 
  &
\beta &= - \alpha q \pm \sqrt{\left(\alpha q\right)^2 - \alpha^2 + 1},
\label{eqn:alpha-beta}
\end{align}
where $q=\xcv_{t}^T\xpv_{t}$, $v=\xpv_{t}^T\xcv^*$, and $w=\xcv_{t}^T\xcv^*$.
Given $\alpha^*$ and $\beta^*$, we can use \refeqn{linear-combo} to update the user vector to $\xcv_{t+1}$.

Recall our goal is to output a sequence of slates. If the preference for $\xp_t$ is positive, we simply use $\xp_t$ to form slate $\xs_t$. If the preference is negative, we do not want to recommend these items; instead, we use the item neighbors of the updated user vector $\sN_{\xx}(\xcv_{t+1}$) to form $\xs_t$. Formally, the slate $\xs_t$ and preference type $p_t$ are defined as: 
\begin{align}
\xs_t &= \begin{cases} 
 \xp_t & \beta > 0 \\
\sN_{\xx}(\xcv_{t+1})  & \beta \leq 0
\end{cases} 
&
p_t &= \begin{cases} 
\textit{more} & \beta > 0 \\ 
\textit{less} & \beta \leq 0. 
\end {cases}
\label{eqn:slate-pref-type}
\end{align}
Along with each $\xs_t$, we also store $p_t$
and item collection $\xp_t$ %
to express the preference for the slate.\footnote{We use an \textit{init} preference type for the first turn of all sequences.} We repeat for $T$ turns per \refalg{seq-gen}.

\begin{algorithm}[t]
\caption{Sequence Generation Procedure}
\label{alg:seq-gen}
\begin{algorithmic}[1]
\renewcommand{\algorithmicrequire}{\textbf{Input:}} 
\renewcommand{\algorithmicensure}{\textbf{Output:}}
\Require
item collections \sZ, item collection types \sT, turns count $T$
\Ensure 
sequence of slates $[(\xs_1, p_1, \xp_1), ..., (\xs_T, p_T, \xp_T)]$
\State $\textrm{seq} = []$
\State $\xcv^* = \textrm{sample}(\sZ)$ \Comment Sample target user vector.
\State $\xcv_1 = \textrm{sample}(\sN_{\xp}(\xcv^*))$ \Comment Sample initial user vector.
\For{$t=1,2,..,T$}
\State $\sigma_t = \textrm{sample}(\sT)$ \Comment Sample collection type. 
\State $\xpv_t = \textrm{sample}(\{\tilde{\bn} \in \sN_{\xp}(\xcv_t) \mid \sigma(\bn) = \sigma_t\})$ \Comment Get collection.
\State $\alpha, \beta = \textrm{findOptimalWeights}(\xcv_t, \xpv_t, \xcv^*)$ \Comment See \refeqns{opt-weights}{alpha-beta}.
\State $\xcv_{t+1} = \alpha \xcv_t + \beta \xpv_{t}$ \Comment Update user vector. 
\State $\xs_t = \textrm{getSlate}(\xc_{t+1}, \xp_t, \beta)$ \Comment See \refeqn{slate-pref-type}.
\State $p_t = \textrm{getPreferenceType}(\beta, t)$ \Comment See \refeqn{slate-pref-type}.
\State $\textrm{seq}.\textrm{append}((\xs_t, p_t, \xp_t))$ \Comment Update slate sequence.
\EndFor
\State return seq
\end{algorithmic}
\end{algorithm}

\Subsection{utt-gen}{User Utterance Generation}

Given a sequence of states, we now generate corresponding user utterances.
Suppose that $\xs_{t-1}$ had ``Running on R\&B'' songs and $\xs_{t}$ ``Cardio Pop'' songs;
the utterance $\xu_t$ should express new preferences (viz., pop music), while assuming prior context (viz., workout music): e.g.,
``How about adding a pop song?''.
Ideal utterances are (1) realistic, using natural language rather than keywords, (2) diverse, covering a wide variety of ways users can express themselves, and (3) contextual, building on previous conversation turns.

We do this using the pretrained dialog inpainter.
First, we manually write system response templates for each preference type and item collection type (e.g., "Of course! Let me add some songs described as <description>. What else?" and "Got it! Let me remove some songs described as <description>. What else?").
We then set up a partial conversation with system responses describing each slate $\xs_t$ by instantiating the above templates with the preference type $p_t$, item collection type $\sigma(\xp_t)$, and item collection metadata $\phi(\xp_t)$.
E.g., to express \emph{more} "Cardio Pop", the system response could be "Of course! Let me add some songs described as \ul{Get those endorphins pumping with this playlist of uptempo pop anthems}. What else?".
Finally, we use a dialog inpainter to generate the ``missing'' user utterances $\xu_t$ (see \reffig{approach} right), completing the output. 
Note that we \emph{only} use the templated responses to prompt the dialog inpainter;
while we only use a small number of prompt templates, the diverse playlist descriptions lead to diverse generated utterances. \emph{User} utterances are entirely produced by inpainting.

\Subsection{case-study}{Case study: Generating a Conversational Music Recommendation Dataset} 

We demonstrate our approach by generating \ourdata, with over one million playlist curation conversations. 

\subsubsection*{\textbf{Input Data}} For collections $\sZ$, we use a proprietary set of expert-curated playlists that consists of theme playlists, where all songs share a common theme (e.g., genre, mood, activity) (19,129 playlists\footnote{Dividing these into train/dev/test (15,276/1,895/1,958) to train the \nonconvoDE.}).
As the manually collected benchmark---CPCD~\cite{chaganty2023cpcd}---contains many queries requesting a specific artist, we also include artist playlists, with all songs sharing one artist (121,704 playlists), derived from the theme playlists.
We call this dataset \emph{\playlistdata}.
$\phi(\xp)$ provides the playlist title and description $\xp$, and $\psi(\xx)$ provides the title, artist names, album, and 128-dimensional audio vector~\cite{mulan-2022} of song $\xx$.

For the \nonconvoDE, we train a dual encoder over \playlistdata{} using a standard contrastive loss~\cite{Oord2018RepresentationLW}. 
Both the queries and items use multi-modal inputs composed of text (WordPiece~\cite{sennrich-etal-2016-neural,kudo-2018-subword}) and audio (MuLan~\cite{mulan-2022}) tokens. The query $q$ is a representation of a playlist $\xp$, generated by concatenating $\phi(\xp)$ with $\psi(\xx)$ for a sample of "seed" songs $\xx \in \xp$.\footnote{We use five seed songs in our experiments.}
The item $x$ is a representation of a randomly sampled song in $\xp$, generated using $\psi(\xx)$. We initialize the dual encoder from a pretrained T5 1.1-Base~\cite{JMLR:v21:20-074} and continue training on the \playlistdata{} theme playlists.

For the dialog inpainter, we use a T5-XXL model pretrained on conversational question-answering and social media discussion threads (similar to InpaintSTOQ~\cite{pmlr-v162-dai22a})---as large-scale conversational recommendation data is not available---and find this works well. 

\subsubsection*{\textbf{Conversation Generation.}} We generate over 1M playlist curation conversations using \ourmethod{}.
We select artist and theme playlist types with equal probability during sequence generation, resulting in user utterances about artists (e.g., ``Can I have some Lady Gaga?'') and broad attributes (e.g. ``How about something more upbeat?''). 
We generate 6 turns per conversation to match CPCD statistics~\cite{chaganty2023cpcd},
then programmatically remove any utterances that:
(1) do not mention the corresponding artists in artist-type queries using string matching,
(2) include offensive language,
(3) are longer than 450 characters, or
(4) have substantial overlap (more than 50 characters) with the following system response. 
We refer to the synthetic dataset as \ourdata{}. \reftab{dataset-stats} shows summary statistics.

\begin{table}[t]
    \centering
    \caption{%
    Summary of \ourdata{} and the manual CPCD\cite{chaganty2023cpcd} dataset;
    Dramatically larger and with similar conversation lengths, \ourdata{} has longer queries and more items per slate.
    We also show the (non-conversational) curated item collection used in this work,  \playlistdata{}.}\label{tab:dataset-stats}%
    \begin{tabular}{l r r r r}
\toprule
& \playlistdata{} & \ourdata{} & \multicolumn{2}{c}{CPCD} \\
\cline{4-5}
Statistic & Train & Train & Dev & Test \\
\midrule
\# of examples & 15,276 & 1,037,701 & 450 & 467 \\
\# of tracks & 332,594 & 332,594 & \multicolumn{2}{c}{106,736} \\
Avg. \# of turns & - & 5.6 & 5.8 & 5.6 \\
Avg. query len. & 106.5 & 80.3 & 53.8 & 55 \\
Avg. \# of items & 53.6 & 47.2 & 19 & 18.3 \\
\bottomrule
\end{tabular}
\end{table}

\subsubsection*{\textbf{Cost Estimates.}}
Querying the \nonconvoDE{} during sequence generation and the dialog inpainter during utterance generation are the most resource-intensive steps of \ourmethod{}.
It takes about 30s to generate each sequence on commodity hardware,
and about 1s to inpaint each conversation using a TPU.
In total, \ourdata{} required about 6,000 vCPU-hours ($\approx$\$200) for sequence generation and about 200 TPUv3-hours ($\approx$\$500) for utterance generation. %
Using spot instances on Google Cloud, this costs about \$700.

\Section{data-quality}{Evaluating Synthetic Data Quality}

We start with intrinsic evaluation by assessing the \emph{quality} of the synthetic dataset using crowdsourcing. 
Note that as the primary use-case for \ourdata{} is training CRSs, it does not need to be indistinguishable from human-collected conversations, 
but should represent \textit{consistent} preferences with corresponding \textit{relevant} slates.\footnote{We show that these criteria are sufficient to train a strong CRS in \refsec{eval}.} %

\begin{table*}[t]
\centering
\caption{
Percentage of results rated as 
Not at all (0), Somewhat (0.5) and Very (1)
in a human evaluation of our synthetic conversations in \ourdata, and human-collected conversations (CPCD).
We also report a weighted average (weights in parentheses).
}
\label{tab:data-eval}
\begin{tabular}{l rrrr l rrrr}
\toprule
\textbf{Question}
& \multicolumn{4}{c}{\textbf{\ourdata{}}} &
& \multicolumn{4}{c}{\textbf{CPCD}}   \\
\cline{2-5} \cline{7-10}
& Not at all & Somewhat & Very & Avg. &
& Not at all & Somewhat & Very & Avg.
\\
\midrule
\textit{How consistent are the preferences?}
& 3.7\%   & 16.5\%  & 79.7\% & 88.0\% &
& 1.0\% & 9.1\% & 89.8\% & 94.4\%
\\
\textit{How relevant is the slate?}
& 4.8\% & 26.2\% & 68.9\% & 82.1\% &
& 1.9\% & 24.1\% & 74.0\% & 86.0\%
\\
\textit{How natural is the conversation?}
& 3.0\% & 46.2\% & 50.8\% & 73.9\% &
& 0.3\%  & 35.3\% & 64.4\% & 82.0\%
\\
\bottomrule
\end{tabular}
\end{table*}

We employed ten skilled annotators to independently review conversations.
They were able to see the full conversation history and tracks in the user playlist, and were asked to use their expert judgment while rating each question in \reftab{data-eval} on a three-point Likert scale.
Specializing in music labeling tasks, they have reasonable pairwise inter-rater agreement for consistency (88.5\%), relevance (81.9\%), and naturalness (72.2\%).

As a point of reference, we compare the ratings to those for human-collected conversations from the Conversational Playlist Creation Dataset~\cite{chaganty2023cpcd} (CPCD).
It consists of music playlist-seeking conversations between two people in a Wizard-of-Oz setting, with one acting as the user, and other acting as a recommender system.
Users came up with a music-listening scenario (e.g., ``a long commute'') and aimed to create a playlist by conversing with the system;
wizards were asked to recommend songs for the user, and could also elicit user preferences.
As \ourdata{} does not include system responses, annotators are shown a uniform system response, ``What else?'', for both datasets.

Annotators reviewed 330 conversations from \ourdata{} and CPCD each, rating 1668 and 1308 turns respectively,\footnote{Turns in CPCD that did not include system responses (e.g., because the system elicited preferences from the user instead) were omitted.}
with each conversation rated by three annotators.
\reftab{data-eval} summarizes the results.
Overall, we find that \ourdata{} contains consistent preferences and relevant slates, close to human-collected conversations in terms of relevance and consistency with a slightly bigger gap in terms of naturalness. 

\subsubsection*{\textbf{How consistent are the user's preferences given the conversation so far?}}
Consistent user preferences are believable and likely given the previous turns (e.g., asking for pop---instead of classical music---when looking for workout music).
Raters found nearly 80\% of turns to be very consistent and fewer than 4\% to be not at all consistent, validating our sequence generation approach.

\subsubsection*{\textbf{How relevant are the results for the user’s preferences given the conversation so far?}} 
System-generated slates should match stated preferences: E.g., if the user asks for romantic songs by Lady Gaga.
Raters found 95.1\% of slates in \ourdata{} at least somewhat relevant, comparable to the expert-selected slates in CPCD.

\subsubsection*{\textbf{How natural do you think this conversation is?}}
Raters evaluating
naturalness were asked: \textit{Can you imagine yourself or someone you know having this conversation?}
We expect lower conversation-level ratings than the turn-level ratings: a single unnatural turn can render a whole conversation unnatural.
Raters find 97\% of conversations in \ourdata{} at least somewhat natural, and 50.8\% very natural.
Interestingly, only 64.4\% of conversations in CPCD were rated very natural despite being human-collected, highlighting how hard it is to guide crowdworkers to create natural conversations.
\Section{eval}{Evaluating Synthetic Data Utility}

Next, we evaluate the \emph{utility} of synthetic data extrinsically on the conversational item retrieval task. 
Using the music domain as an example, we show that \ourdata{} can be used to train a CRS that significantly outperforms both unsupervised baselines as well as systems trained on expert-curated playlists.

\Subsection{exp-setup}{Experimental Setup}

We focus on the task of conversational item retrieval to show the utility of synthetic data to create such a system. Utility could also be measured for other conversational applications like preference elicitation, but we omit these due to space constraints.

\subsubsection{Formalism.}

Assume
a CRS where the user has a target set of items (\emph{slate}) $\xs^{*}$ in mind,\footnote{%
    To simplify evaluation, we assume the user's target is fixed during a conversation.
} and in each turn they provide a natural language statement (query) $\xu_t$. In turn the system responds with a slate $\xs_t$ to satisfy the user's query in light of the conversation history. Formally, given query $\xu_t$ and conversation history, $\bH_t = \left(\xu_1, \xs_1, \dots, \xu_{t-1}, \xs_{t-1}\right)$, the CRS predicts a new slate of items that ranks the user's target $\xs^*$ highest.

\subsubsection{Benchmark Dataset.}

We use the CPCD dataset~\cite{chaganty2023cpcd} to evaluate recommendation performance.
Note that each conversation has multiple turns, each having a user query and result slate, with binary like/dislike ratings for each item.
As we do not model explicit user feedback here, we remove disliked songs from slates in the conversation history.
The target slate $\xs^*$ thus includes liked songs across all turns of the conversation.
Systems are evaluated on their ability to retrieve songs in this target from a corpus of 106k songs given the user query and conversation history in each turn.

\subsubsection{Evaluation Metrics.} 
\label{sec:eval-metrics}
Following %
the setup of CPCD~\cite{chaganty2023cpcd}, 
we compare systems using a standard ranking metric, Hits@$k$,
which is 1 if and only if any of the top-k retrieved items are in the target slate, and report performance at several values of $k$ (10, 20, 100).  
Unless otherwise stated, we report macro-averaged Hits@$k$, averaging Hits@$k$ across turns within a conversation and then across conversations.
To support evaluation over multiple turns, we must take into account several considerations.
First, the target slate at each turn only consists of songs that have not been seen in the history up to that point. 
\emph{As the metrics depend on the target slate---which changes across turns---we cannot directly make performance comparisons across turns.}
To compare across models \emph{within} a turn, we use the same history and target slate for all models, assuming a ``gold'' history (rather than building the history using model predictions). 
Second, at each turn, songs may be liked by users, but not added to the history (e.g., when there is a limit to the number of songs in the history).
Following \cite{paraparleftovers}, we refer to these songs as leftovers and keep them in the target slate, resulting in a larger target slate compared to removing all previously liked songs.

\Subsection{model}{Training using Synthetic Data} %

We now show how our synthetic dataset, \ourdata, can be used to train a conversational music recommender using synthetic data only (\oursystem{} and \oursystemcr{}) as well as a combination of real and synthetic data (\oursystemcr$\rhd$CPCD). 

\Subsubsection{model-inputs}{Model}

We model the CRS as \textit{a ranking function} $\rho: \sX \to \R$ using a dual encoder~\citep{gillick-etal-2019-learning,dpr,gtr} as they are 
recognized as particularly effective recommendation algorithms. They independently embed queries $q \in \sQ$ and items $x \in \sX$ into normalized dense vectors,
and rank items using cosine similarity.
At turn $t$, we use the conversation history $H_t$ and the latest user utterance $\xu_t$ to construct a query $q_t$, and then 
predict a slate $\xs_t$ whose items $x$ maximize the ranking function $\rho(x; q_t) = f(q_t)^\top g(x)$, i.e., are closest to $q_t$ in embedding space.
We construct $q_t$ by concatenating the utterances with text representations of the top-k songs in each slate in reverse chronological order.\footnote{In our experiments, we use up to three songs from each slate.}
For example, the query at turn $t$ is: 
\begin{equation*}
\resizebox{0.7\hsize}{!}{$u_t$ \sep d($\xs_{t-1}$) \sep $u_{t-1}$ \dots~d($\xs_1$) \sep $u_1$}
\end{equation*}
where $d(\xs_t)$ is a textual representation of $\xs_t$, 
and \sep represents a separator token.
We use the same text representation for items as in the \nonconvoDE{} (\refsec{case-study}).

\Subsubsection{pretraining_and_inference}{Training and Inference}
\label{pretraining_and_inference}

We initialize our dual encoder from a pretrained T5 1.1-Base~\citep{JMLR:v21:20-074} and train on \ourdata{} for 100k steps using Adafactor~\cite{Shazeer2018-pm} with batch size 512 and constant learning rate 1e-3.
We use a standard contrastive loss with in-batch negatives:
given example $i$ with query $q_i$ and target slate $\xs_i$,
we randomly sample an item $x_i \in \xs_i$ to be a positive target,
and take items from other slates in the batch, $x_j$ where $j \neq i$, to be negative targets.
To improve robustness, we augment our training data as follows:
(1) we generate conversations of varying lengths by randomly truncating conversations to its first $t$ turns, and
(2) we generate slates of varying lengths by randomly truncating slates to their first $k$ items.
We write \oursystem{} to denote the standard dual encoder (DE) trained on \ourdata{}.
We select the checkpoint with %
highest Hits@10 on the CPCD development set from steps 25k, 50k, 75k, and 100k.
We also consider three Contriever-based variants \citep{izacard2021contriever} by fine-tuning an existing Contriever checkpoint. For \oursystemcr{}, we follow Contriever's training procedure and use randomly extracted spans from the query and the target respectively to form a positive pair. We also use the same hyper-parameters that are used for training the base Contriever checkpoint whenever possible. The model is trained only on synthetic data for 1000 steps with a learning rate of 1e-3. \oursystemcr$\rhd$CPCD is further fine-tuned on real data, i.e., CPCD. We train \oursystemcr$\rhd$CPCD for 100 steps as CPCD is a small dataset, and we use a batch size of 64 and a learning rate of 1e-4. CR$\rhd$CPCD is directly fine-tuned from the base Contriever checkpoint with CPCD for 100 steps using a batch size of 64 and a learning rate of 1e-4.

For inference, we build an index of pre-computed item embeddings.
We embed queries as in training, and use nearest neighbor search to return a slate $\xs_t$ with the top-k items for $q_t$.

\Subsection{baselines}{Baselines}

We compare against three baselines that have not been trained on the target dataset:
(1) BM25, a sparse retrieval, bag-of-words baseline;
(2) Contriever (Cr)~\citep{izacard2021contriever}, an unsupervised dense retrieval baseline; and
(3) \nonconvo{}, a dense retriever trained over playlist description-song pairs from \playlistdata{} (i.e., the \nonconvoDE{} in \refsec{case-study}).
We also consider a fourth, fine-tuned variant, Cr$\rhd$CPCD: Contriever fine-tuned on the CPCD dataset.\footnote{Our results are better than those presented in \citep{chaganty2023cpcd} due to using a smaller batch size of 64, a learning rate of 1e-4, and training less steps (100) to avoid overfitting.}

Our BM25 system includes conversation history by concatenating previous user queries (omitting conversation history degrades performance).
Contriever is implemented using a T5-Base architecture and trained on span pairs from the C4 dataset following \cite{izacard2021contriever}.
\nonconvo{} is implemented identically to \oursystem{}, but using a different training dataset (\playlistdata{});
at test time, we use the same conversation history including previous queries and results.

\Subsection{model-results}{Results} 
Our key result from the comparison of models in \reftab{benchmark_results} is that simply training a standard dual encoder on \ourdata{} suffices to build a strong music CRS that outperforms baselines that have not been trained on the target dataset (\oursystem{} vs. BM25, \nonconvo{}, Contriever).
Of these baselines, BM25 performs best. We attribute its stronger performance over \nonconvo{} and Contriever to the prevalence of artist and song-specific queries in CPCD. 
On these queries, matching keywords suffices, and bag-of-words-based models can perform well. 

We also find that training on synthetic data can outperform a state-of-the-art baseline trained on the target dataset in terms of Hits@$100$ (Cr$\rhd$CPCD vs. \oursystemcr).
A combination of synthetic and real data (\oursystemcr$\rhd$CPCD) yields further performance improvements.

\begin{table}[t]
\centering
\caption{\label{tab:benchmark_results} CPCD test set results. Highest scores are boldfaced.} %
\begin{tabular}{lrrr}
\toprule
Model                                     & Hits@10  & Hits@20  & Hits@100 \\
\midrule
BM25                  & {19.7} & {27.4} & {45.5} \\
\midrule
DE$\rhd$\playlistdata{}
                      & {13.1} & {19.6} & {43.2} \\
\sysname              & 20.8 & 31.1 & 52.1 \\
\midrule
Contriever (Cr)          & {16.2} & {23.1} & {39.4} \\
Cr$\rhd$CPCD & \bf{32.6} & 42.0 & 58.0 \\
\oursystemcr$^*$ & 29.9 & 41.0 & 60.7 \\
\oursystemcr$^*\rhd$CPCD & 30.6 & \bf{42.5} & \bf{62.1} \\
\bottomrule
\multicolumn{4}{p{8cm}}{\footnotesize $^*$ Following the Contriever training procedure~\citep{izacard2021contriever}, we extract spans from the data, as detailed in Section~\ref{pretraining_and_inference}.}
\end{tabular}

\end{table}

\begin{table}[t]
\caption{%
    Example conversations from an online human evaluation. The top-ranked song from both systems is shown under the query, with
    \cmark and \xmark{} denoting ratings;
    \music{} symbols link to the song online.
    Unlike BM25, \oursystem{} retrieves relevant songs even when there is no literal match (``Happiness'') and maintains context across turns (``Mother'').
}
\label{tab:end-to-end-examples}
\begin{tabularx}{\columnwidth}{l p{0.808\columnwidth}}
\toprule
\emph{User} & 
\emph{Can you make me a playlist of sad songs, I broke up with girlfriend} \\
\shortoursystem{}\!\!\!
 & \cmark{} Happiness by Hobo Johnson \music{0QoTwItJM_I} \\
BM25
 & \xmark{} I Can't Make You Love Me by Teddy Swims \music{OwCD9svuUU0} \\
\emph{User} & \emph{can you add more of female voices} \\
\shortoursystem{}\!\!\!
 & \cmark{} Mother by Courtney Love \& The Turning \music{5IHTLP6gx-U} \\
BM25
 & \xmark{} I Can't Make You Love Me by Dave Thomas Junior \music{3RPzrEopxDc} \\
\midrule
\emph{User} & 
\emph{Looking to create a playlist to help me sleep. Soft sounds with minimal lyrics}. \\
\shortoursystem{}\!\!\!
& \cmark{} Suburban Call For Concrete Dreams by Ave Air
     \music{40VAWQmSq9k} \\
BM25
& \cmark{} Soul Blind With Lyrics by Shane Phipps 
    \music{AFLljlS1iW8} \\
\emph{User} & 
\emph{Piano music please.
Maybe add some classical music.} \\
\shortoursystem{}\!\!\!
& \cmark{} Rise by Arelius \music{DIfIEC65yKw} \\
BM25
& \xmark{} Morning Mood - Grieg - Classical Piano - Classical Sleep Music and Ocean Sounds\dots \music{Pjv9I8TB03g} \\
\bottomrule
\end{tabularx}
\end{table}

\Subsection{result-analysis}{Result Analysis} 
We now validate our offline results via an online experiment, performing additional analysis to better understand what contributes to our model's performance.
Due to space constraints, we focus on the standard dual encoder trained on synthetic data (\oursystem{}).

\subsubsection*{\textbf{Do offline results agree with online evaluation?}} %

\begin{figure}[t]
    \centering
    \includegraphics[width=0.7\columnwidth,trim=0 1cm 0 0]{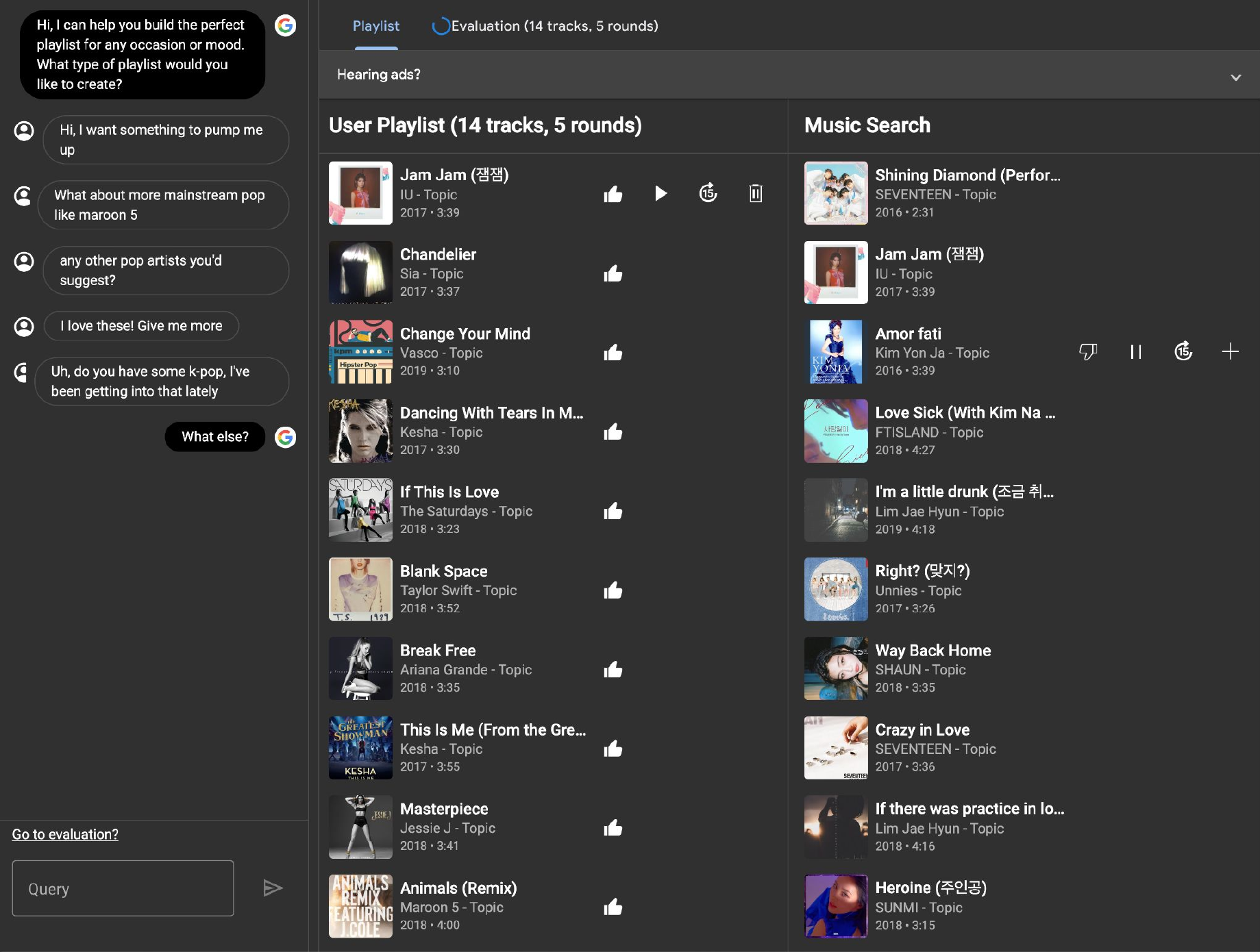}
    \caption{%
    Online evaluation system screenshot, showing past queries, rated songs and retrieved slate.
    }
    \label{fig:end-to-end-interface}
\end{figure}

It is well known that users interact with online systems differently based on their results, and may rank systems differently than in offline evaluation~\cite{cis-survey,Jannach2022-ow}.
We thus conduct an online evaluation to compare the best performing baseline (BM25) against \oursystem{}.

We recruited 11 fluent English speakers, based in the United States, who regularly listen to music from an online crowdworking platform.
Users interacted with a web-based interface (\reffig{end-to-end-interface}) that allowed them to express their preferences conversationally using any language they chose.
Users were provided instructions and %
an overview of the interface and its features.
Similar to \citep{chaganty2023cpcd}, they were asked to start each conversation with a %
context for their listening session, e.g., ``I want to create a playlist to pump me up when I'm feeling tired,'' and encouraged to express broad and general preferences, but were otherwise not constrained in their interaction with the system.
After each turn, users were required to rate all the songs in the result slate and to complete at least 5 rounds of conversation before proceeding to evaluation.

Finally, to ensure the same user does not rate songs differently based on their conversation history or their previous experience with the system, we present raters a single randomized combined slate of results that ensures a lack of bias towards either system using \textit{team-draft interleaving}~\citep{Radlinski2008-hl}, a standard combination method (algorithm details omitted here due to space limitations).

We collected 227 conversations and 2454 item ratings after filtering out conversations with an average utterance length less than four, and any turns with irrelevant utterances like ``hello'' or ``thank you.''
We then computed per-turn hit rates for the two systems, finding that \oursystem{} had significantly higher Hits@10 (69.8\%) than BM25 (62.0\%) with $p < 0.01$, which aligns with the offline evaluation results (cf. Table~\ref{tab:benchmark_results}).
Qualitatively, \oursystem{} was also better able to understand broad queries and refinements and returned non-literal results to the user while BM25 tended to overemphasize lexical overlap (\reftab{end-to-end-examples}).

\subsubsection*{\textbf{How do models compare across turns?}}
\begin{figure}[t]
	\centering
	\includegraphics[width=0.8\columnwidth,trim=0 1.5cm 0 0]{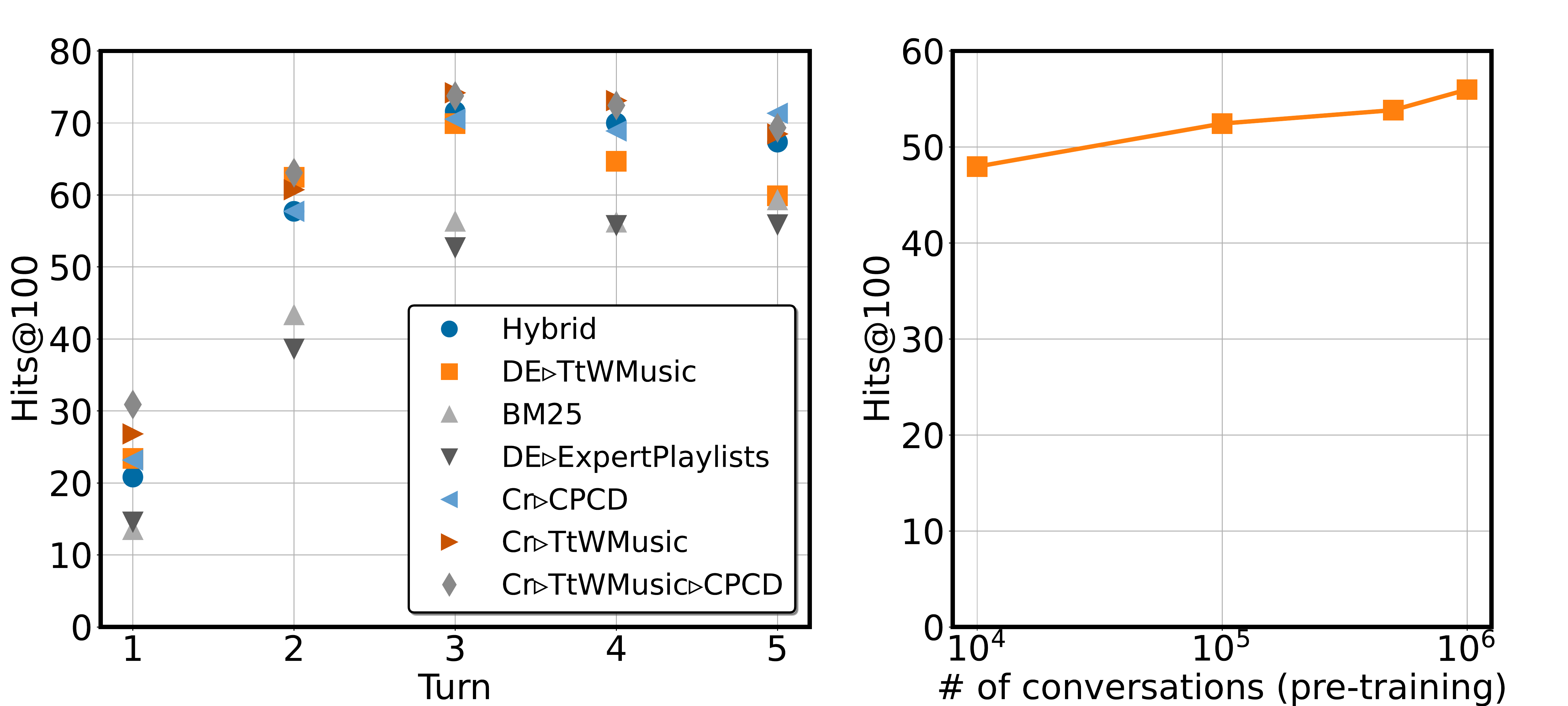}
\caption{
Hits@100 on CPCD test.
\textit{(Left)}
Results (averaged) per conversation turn. %
\textit{(Right)} \oursystem{} performance as a function of training data size.
}
\label{fig:plots}
\end{figure}

Figure~\ref{fig:plots} (left) compares model turn-level performance. %
We show the first five turns, averaging Hits@100 across conversations for each turn. 
Note that as conversations vary in length, the number of conversations in each turn vary (down to 349 conversations at Turn 5).
Also note that the distribution of queries in CPCD change as conversations progress \cite{chaganty2023cpcd}.
The target slate size also decreases as songs are included in the conversation history; thus, we cannot compare across turns. 
We observe all models that used \ourdata{} for pre-training outperform BM25 and \nonconvo{} across turns, suggesting \ourdata's effectiveness.

\subsubsection*{\textbf{How does performance scale with the amount of training data?}} 
Figure~\ref{fig:plots} (right) shows Hits@100 on CPCD test as we vary the number of conversations in the synthetic training dataset. We find that the performance improves from 48.0\% to 56.0\% for 10k to 1M conversations and does not level off, suggesting that generating more conversations may further improve performance. In contrast, most crowdsourced or real-world conversational recommendation datasets contain <30k conversations (e.g., \citep{li2018towards, moon-etal-2019-opendialkg, zhou-etal-2020-towards,liu-etal-2021-durecdial,taskmaster2,jia-etal-2022-e}).

\begin{table}[t]
\caption{
\label{tab:ablations}
Synthetic data ablations. We report results for DE$\rhd$TtWMusic using 100k conversations.
}
\begin{tabular}{lrrr}
\toprule
Training Dataset 
& Hits@10  & Hits@20  & Hits@100 \\
\midrule
\ourdata-100k                 & \bf{20.3} & \bf{30.2} & \bf{52.4} \\
$-$ Sequence Gen. & 11.9 & 16.3 & 28.9 \\
$-$ Utterance Gen. & 18.9 & 26.7 & 43.4 \\
\bottomrule
\end{tabular}
\end{table}

\subsubsection*{\textbf{How important are the synthetic data generation components?}} 
We evaluate the importance of both components in \ourmethod{}: %
To ablate sequence generation, we use a random sequence generator which randomly selects an item collection $\xp_t$ each turn to use as the slate (i.e., $\xs_t = \xp_t$).
We then use our standard method to generate utterances with dialog inpainting. 
As this method does not encourage consistency between turns, we expect the synthetic data to be less useful as training data. 
To ablate dialog inpainting, we use our standard method for generating sequences of slates.
Then, instead of using dialog inpainting, we use templated user utterances, similar to the templated system responses. %
This generates less diverse and realistic user utterances, so we also expect the synthetic data to be less useful as training data.  
Table~\ref{tab:ablations} shows the results on CPCD test when we train models on 100k conversations in each dataset.\footnote{We train on smaller datasets for ablations than the main results (100k vs $>$1M conversations) to limit computational cost, as we assessed multiple ablations. Others are not reported due to space constraints.}
We see that removing either component leads to significant drops in Hits@$k$ at all values of $k$.%

\Section{conclusion}{Conclusion}

We introduced a general technique to convert curated item collections into synthetic item set curation conversations, and demonstrated the benefits of building a conversational recommendation system using such data in the music domain.
\ourmethod{} can be easily adapted to other domains and other languages given corresponding item collections and an appropriate language model.

While our work has focused on synthetic data generation in the context of a specific use-case, it holds significant implications for future applications across various domains. The ability to efficiently generate synthetic data helps address the challenge of data scarcity, as well as important considerations of provenance of user data with respect to privacy when constructing datasets. It also offers a way to study novel tasks for which training data at scale is unavailable. Within this context, the evaluation methodology employed in this paper, namely, assessing synthetic data intrinsically against real data and extrinsically by leveraging it in a downstream application, is applicable in a broad range of scenarios.

Finally, noting that language models are used to create the conversational utterances, further work is warranted assessing what biases may be present in the utterances produced~\citep{foundation-models}, and how they can be reduced.

\section*{Ethical Considerations}
We now briefly consider key ethical considerations. 

First, it is important to consider the provenance of the user generated content that is provided as input to synthetic data generation, to establish any constraints on how the synthetic data may be used. In the case of playlists and associated metadata used here, we restrict ourselves to the ExpertPlaylist dataset described, for which an ethical review established the data can be used for the purposes of this research. Similarly, it is important to consider the properties of the dialog inpainting model, in particular considering how utterance generation is influenced by that model's training data.

Beyond this, our conclusions are based on the judgments of a relatively small pool of crowdworkers. While they were knowledgeable in the target domain and were trained appropriately for our tasks, there is a potential risk that their conclusions may not be representative of the broader population. In such a case, the conclusions might not generalized. In particular, we note that all experiments in this work were performed using English-language data. Hence it is unknown whether the conclusions would match for people from other cultural and linguistic backgrounds.

Finally, we note that all raters used in this work are paid contractors. They received their standard contracted wage, which is above the living wage in their country of employment.

\bibliographystyle{ACM-Reference-Format}
\bibliography{custom}

\end{document}